\documentclass[12pt]{iopart}
\usepackage{graphics}
\begin{document}



\title{CMS search plans and sensitivity to new physics with dijets}
\author{Anwar Bhatti$^1$, Benjamin Bollen$^2$, Marco Cardaci$^2$, 
Frank Chlebana$^3$, Selda Esen$^4$, Robert M. Harris$^3$, Manoj K. Jha$^5$, 
Konstantinos Kousouris$^3$, David Mason$^3$ and Marek Zielinski$^6$}
\address{$^1$ Rockefeller University, New York, NY, USA}
\address{$^2$ Universiteit Antwerpen, Antwerp, Belgium}
\address{$^3$ Fermilab, Batavia, IL, USA}
\address{$^4$ Brown University, Providence, RI, USA}
\address{$^5$ University of Delhi, India and INFN, Bologna, Italy }
\address{$^6$ University of  Rochester, Rochester, NY, USA} 
\ead{rharris@fnal.gov}

\begin{abstract}
CMS will use dijets to search for physics beyond the standard model during early LHC running.
The inclusive jet cross section as a function of jet transverse momentum, 
with 10 pb$^{-1}$ of integrated luminosity, is sensitive 
to contact interactions beyond the reach of the Tevatron. The dijet mass
distribution will be used to search for dijet resonances coming from new particles,
for example an excited quark. Additional sensitivity to the existence of contact interactions 
or dijet resonances can be obtained by comparing dijet rates in two distinct 
pseudorapidity regions.
\end{abstract}

\pacs{12.38.Qk, 12.60.Rc, 13.87.Ce}
\submitto{\jpg}
\maketitle

The Large Hadron Collider at CERN will produce many events 
with two energetic jets resulting from proton-proton collisions at $\sqrt{s}=14$ TeV. 
These dijet events result from parton scattering, produced by  
the strong interaction of quarks ($q$) and gluons ($g$) inside the protons. This paper 
discusses plans to use dijets in the search for two signals of new physics: contact 
interactions and resonances decaying into dijets.  Two models of quark compositeness 
have been considered for this generic search.
The first model is a contact interaction~\cite{refELP} among left-handed quarks 
at an energy scale $\Lambda^+$ in the process $qq\rightarrow qq$, modeled with the 
effective Lagrangian 
$L_{qq} = (\pm 2\pi/\Lambda^2) (\overline{q}_{L} \gamma^{\mu}q_L) (\overline{q}_L
\gamma_\mu q_L)$ with $+$ chosen for the sign.
The second is a model of an excited quark ($q$*)~\cite{refBaur} in the 
process $qg\rightarrow q* \rightarrow qg$,
detectable as a dijet resonance. All processes presented here have been simulated 
using PYTHIA version 6.4~\cite{refPYTHIA}.

A detailed description of the Compact Muon Solenoid (CMS) experiment can be found 
elsewhere~\cite{refCMS,refPTDR1}. 
The CMS coordinate system has the origin at the center of the detector, 
$z$-axis points along the beam direction toward the west, with the transverse 
plane perpendicular to the beam. We define $\phi$ to be the azimuthal 
angle, $\theta$ to be the polar angle and the pseudorapidity as 
$\eta=-\ln(\tan[\theta/2])$. 
The central feature of the CMS apparatus is a superconducting solenoid, of 6 m 
internal diameter. 
Within the field volume are the silicon pixel and strip tracker, and the barrel and
endcap calorimeters ($|\eta|<3$): a crystal electromagnetic calorimeter (ECAL) and
a brass-scintillator hadronic calorimeter (HCAL).
Outside the field volume, in the forward region, there is 
an iron-quartz fiber hadronic calorimeter ($3<|\eta|<5$).
The HCAL and ECAL cells are grouped into towers, projecting radially outward from 
the origin, for triggering purposes and to facilitate the jet reconstruction.  
In the region 
$|\eta|<1.74$ these projective calorimeter towers have segmentation 
$\Delta\eta = \Delta\phi = 0.087$, and the $\eta$ and $\phi$ width 
progressively increases at higher values of $\eta$. 
The energy in the HCAL and 
ECAL within each projective tower is summed to find the calorimeter tower energy.
Towers with $|\eta|<1.3$ contain only cells from the barrel 
calorimeters, towers in the transition region $1.3<|\eta|<1.5$ contain a mixture of 
barrel and endcap cells, and towers in the region $1.5<|\eta|<3.0$ 
contain only cells from the endcap calorimeters. 

Jets are reconstructed using both the iterative and midpoint cone 
algorithms~\cite{refPTDR1}, with indistinguishable results for this analysis.
Below we will discuss three types of jets: reconstructed, corrected and generated.
The reconstructed jet energy, $E$, is defined as the scalar sum of the calorimeter tower 
energies inside a cone of radius 
$\sqrt{(\Delta\eta)^2 + (\Delta\phi)^2}=0.5$, centered on the jet axis. 
The jet momentum, $\vec{p}$, is the corresponding vector sum: 
$\vec{p} = \sum{E_i\hat{u}_i}$ with $\hat{u}_i$ being the 
unit vector pointing from the origin to the energy 
deposition $E_i$ inside the cone. The jet transverse momentum, $p_T$, is the component 
of $\vec{p}$ in the transverse plane. 
The $E$ and $\vec{p}$ of a reconstructed jet are then corrected for the  
non-linear response of the calorimeter to a generated jet. 
Generated jets come 
from applying the same jet algorithm to the Lorentz vectors of stable generated particles
before detector simulation.
On average, the $p_T$ of a corrected jet is equal 
to the $p_T$ of the corresponding generated jet.
The corrections estimated from a GEANT~\cite{refGEANT} simulation of the CMS detector increase 
the average jet $p_T$ by roughly
50\% (10\%) for 70 GeV (3 TeV) jets in the region $|\eta|<1.3$. 
The applied corrections depend on jet $\eta$ as well as $p_T$.
The jet measurements presented here are within the region $|\eta|<1.3$,
where the sensitivity to new physics is expected to be the highest,
and where the reconstructed jet response variations as a function of $\eta$ 
are both moderate and smooth.
Further details on jet reconstruction and jet energy corrections can be found
elsewhere~\cite{refPTDR1,refJetAlgPAS}. 

The dijet system is composed of the 
two jets with the highest $p_T$ in an event (leading jets), 
and the dijet mass is given by 
$m=\sqrt{(E_1 + E_2)^2 - (\vec{p}_1 + \vec{p}_2)^2}$.  
The estimated dijet mass resolution varies from 9\% at a dijet mass 
of 0.7 TeV to 4.5\% at 5 TeV.

CMS will record events that pass a first level trigger followed by a high level 
trigger. 
For an instantaneous luminosity of $10^{32}$ cm$^{-2}$s$^{-1}$, we consider three event samples  
collected by requiring at least one jet in the high level trigger with 
corrected transverse energy 
above 60, 120 and 250 GeV, prescaled by factors of 2000, 40 and 1, respectively.  
For an integrated luminosity of 100~pb$^{-1}$, the three event samples will effectively correspond to 
0.05, 2.5, and 100~pb$^{-1}$.
The first event sample will be used to measure the trigger efficiency 
of the second sample.
The second and third event samples will be used to study dijets of mass above 330 and 
670 GeV, respectively, for which the trigger efficiencies  
are expected to be higher than 99\%~\cite{refPTDRTrigger}. 

Backgrounds from cosmic rays, beam halo, and detector noise are 
expected to occasionally produce events with large or unbalanced energy depositions. They 
will be removed by requiring \mbox{${\not\!\!E_T}/\sum E_T < 0.3$} and 
$\sum E_T<14$ TeV, where  \mbox{${\not\!\!E_T}$} ($\sum E_T$) is the magnitude of the vector (scalar)
sum of the transverse energies measured by all calorimeter towers in the event. This cut is estimated 
to be more than 99\% efficient for both QCD jet events and the signals
of new physics considered. In the high $p_T$ region relevant for this search, 
jet reconstruction is fully efficient.

\begin{figure}[hbt]
  \begin{center}
       \resizebox{7in}{!}{\includegraphics{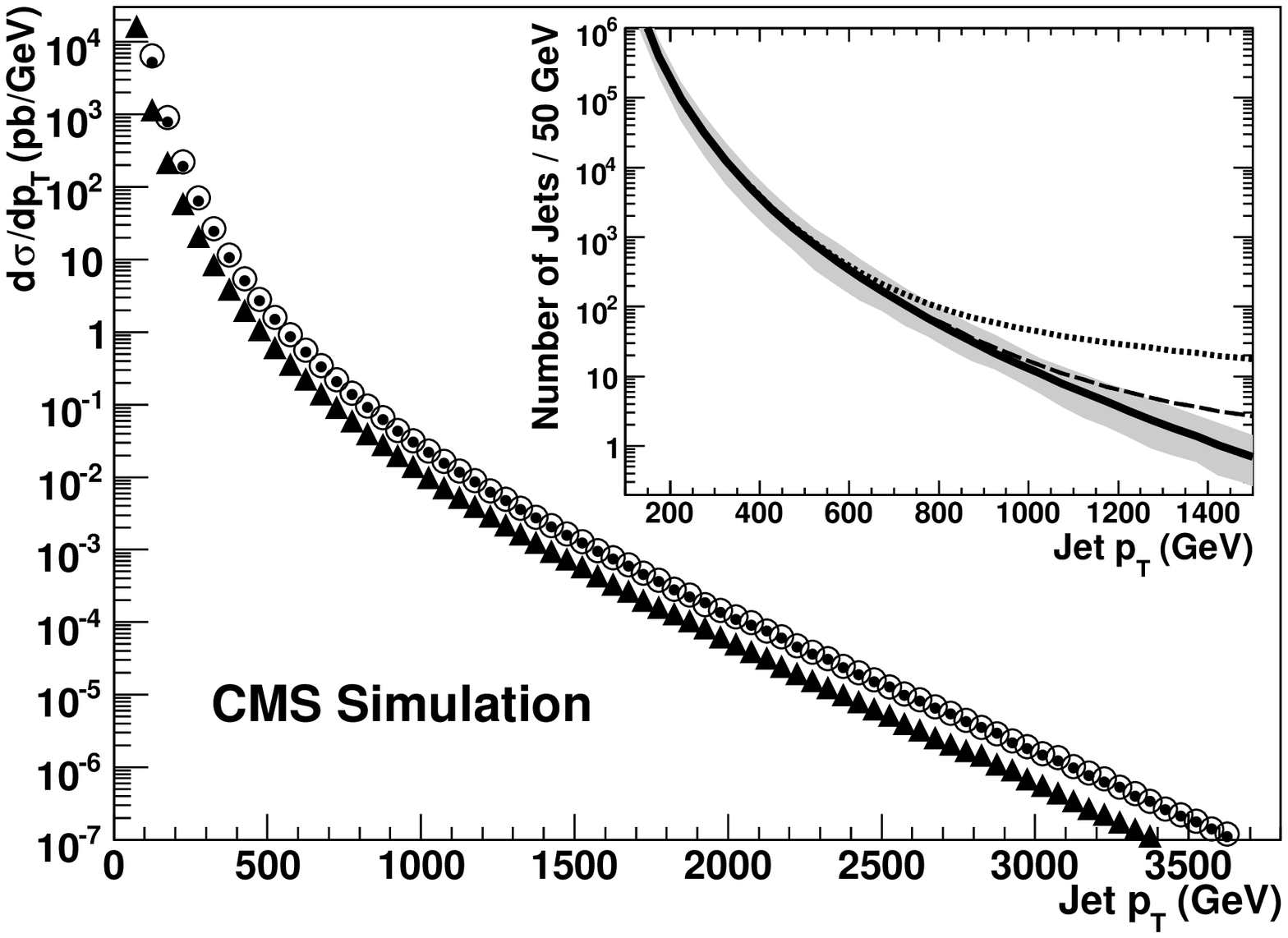}}
    \caption{ The inclusive jet $p_T$ differential cross section expected from QCD 
for $|\eta|<1$, for generated jets (points), reconstructed jets (triangles), and corrected 
jets (open circles). The inset shows the number of generated jets expected  
in 50 GeV bins for an integrated luminosity of 10 pb$^{-1}$.
The standard QCD curve (solid) is modified by a signal 
from contact interactions with scale $\Lambda^+=3$ TeV (dotted) and 5 TeV (dashed). The
shaded band represents the effect of a 10\% uncertainty on the jet energy scale.}
    \label{fig1}
  \end{center}
\end{figure}

\begin{figure}[hbt]
  \begin{center}
       \resizebox{7in}{!}{\includegraphics{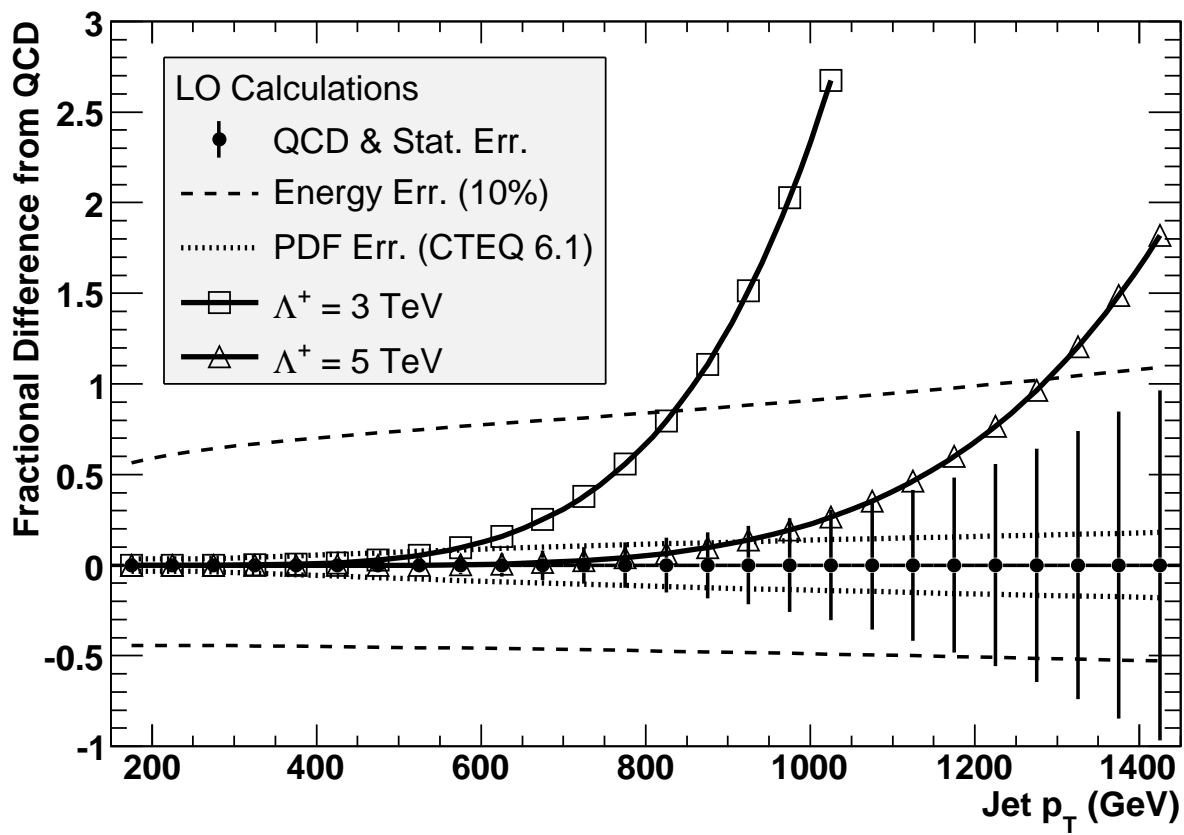}}
    \caption{ The fractional difference from the QCD jet rate resulting from 
a 10\% uncertainty on the jet energy scale (dashed), uncertainties in parton 
distributions (dotted), and signals from contact interactions with scale 
$\Lambda^+=3$ TeV (boxes) and $\Lambda^+=5$ TeV (triangles).  Statistical 
uncertainties expected for an integrated luminosity of 10 pb$^{-1}$ (vertical 
bars) are shown on the QCD prediction (points).}
    \label{fig2}
  \end{center}
\end{figure}

CMS plans to search for contact interactions using the jet $p_T$ distribution. 
Figure~\ref{fig1} shows simulations of the inclusive jet differential cross section as a function
of $p_T$, for jets with $|\eta|<1$. 
Considering first the QCD processes, the reconstructed and corrected 
quantities are compared with the QCD prediction for generated jets. After corrections, the reconstructed and 
generated distributions agree. The ratio of the corrected
jet cross section to the generated jet cross section varies between 1.2 at 
$p_T$ = 100 GeV and 1.05 at $p_T$ = 500 GeV, remaining roughly constant for higher $p_T$. 
The deviation of this ratio from 1 is attributed to the smearing effect of the jet $p_T$ resolution 
on the steeply falling spectrum.
The measured spectrum in data could be further corrected for 
resolution smearing, and this ratio from simulation is an estimate of
the size of that correction. The measurement uncertainties are predominantly systematic.  
The inset in Fig.~\ref{fig1} shows the effect on the jet rate of a 10\% uncertainty  
in the jet energy correction.  
Fig.~\ref{fig2} also shows the effect of this uncertainty on a lowest order QCD calculation. 
This level of jet energy uncertainty could be expected in
early running, for an integrated luminosity around 10 pb$^{-1}$. This experimental uncertainty 
is roughly an order of magnitude larger than the uncertainties from parton 
distributions, as estimated using CTEQ6.1 fits~\cite{refCTEQ} and shown in Fig.~\ref{fig2}.
Figures~\ref{fig1} and~\ref{fig2} show that the effect 
of new physics from a contact interaction with scale $\Lambda^{+}=3$ TeV is convincingly above what could be expected 
for measurement uncertainties with only 10 pb$^{-1}$. 
For comparison, a Tevatron search has excluded contact interactions with scales $\Lambda^{+}$ 
below 2.7~TeV~\cite{refD0}. The results of the lowest order calculations in 
Fig.~\ref{fig2} are the same as the simulation results in the inset to Fig.~\ref{fig1}.

\begin{figure}[hbt]
  \begin{center}
       \resizebox{7in}{!}{\includegraphics{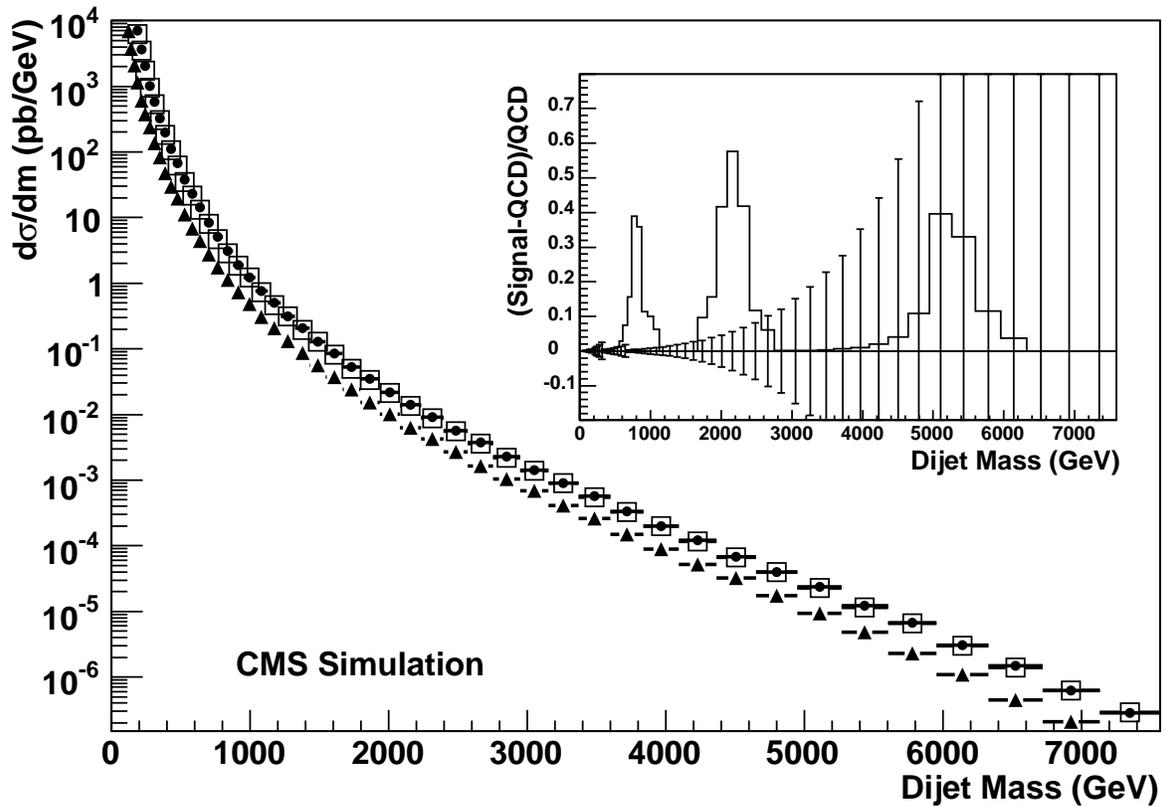}}
    \caption{ 
The dijet mass differential cross section expected from QCD for $|\eta|<1$ 
from generated jets (points), reconstructed jets (triangles), and corrected 
jets (open boxes).  The inset shows dijet resonances reconstructed using 
corrected jets, coming from $q$* signals~\cite{refPTDRresonances} of mass $0.7$, $2$, and $5$ TeV.
The fractional difference (histogram) between the $q$* signal and the QCD background 
is compared to the statistical uncertainties in the QCD prediction (vertical bars) for an integrated 
luminosity of 100 pb$^{-1}$.}
    \label{fig3}
  \end{center}
\end{figure}

CMS plans to search for narrow dijet resonances using the dijet mass distribution.
Figure~\ref{fig3} shows the differential cross section versus 
dijet mass, where both leading jets have $|\eta|<1$, and the mass bins have a width roughly equal to 
the dijet mass resolution.  Considering first the QCD processes, the 
cross section for corrected jets agrees with the QCD prediction 
from generated jets. To determine the background shape either the Monte Carlo prediction or 
a parameterized fit to the data can be used.  The inset to Fig.~\ref{fig3} shows a 
simulation of narrow dijet resonances with a $q$* production cross 
section. For $q$* masses of $0.7$, $2.0$ and $5.0$ TeV the cross sections for jet 
$|\eta|<1$ are $795$, $9.01$ and $0.0182$ pb, respectively. This is
compared to the statistical uncertainties in the QCD prediction, including trigger prescaling.
This comparison shows that with an integrated luminosity of 100 pb$^{-1}$ a $q$* dijet resonance with a mass
of 2 TeV would produce a convincing signal above the statistical uncertainties 
from the QCD background.
For comparison, a Tevatron search has excluded $q$* dijet resonances with 
mass, M, below $0.87$ TeV~\cite{refCDFrun2}. 
The heaviest dijet resonances that CMS can discover (at five standard deviations) with 100~pb$^{-1}$ 
of integrated luminosity, using this search technique and including the expected
systematic uncertainties~\cite{refPTDR2,refPTDRresonances}, are: 2.5~TeV for $q$*, 
2.2~TeV for axigluons~\cite{refAxi} or colorons~\cite{refColoron}, 2.0~TeV for 
$E_6$ diquarks~\cite{refDiq}, and 1.5~TeV for color octet technirhos~\cite{refTrho}.
Studies of the jet $\eta$ cut have concluded that the optimal sensitivity to new physics is 
achieved with $|\eta|<1.3$ for a 2 TeV spin 1 dijet resonance decaying to $q\bar{q}$~\cite{refPAS}.

\begin{figure}[hbt]
  \begin{center}
      \resizebox{7in}{!}{\includegraphics{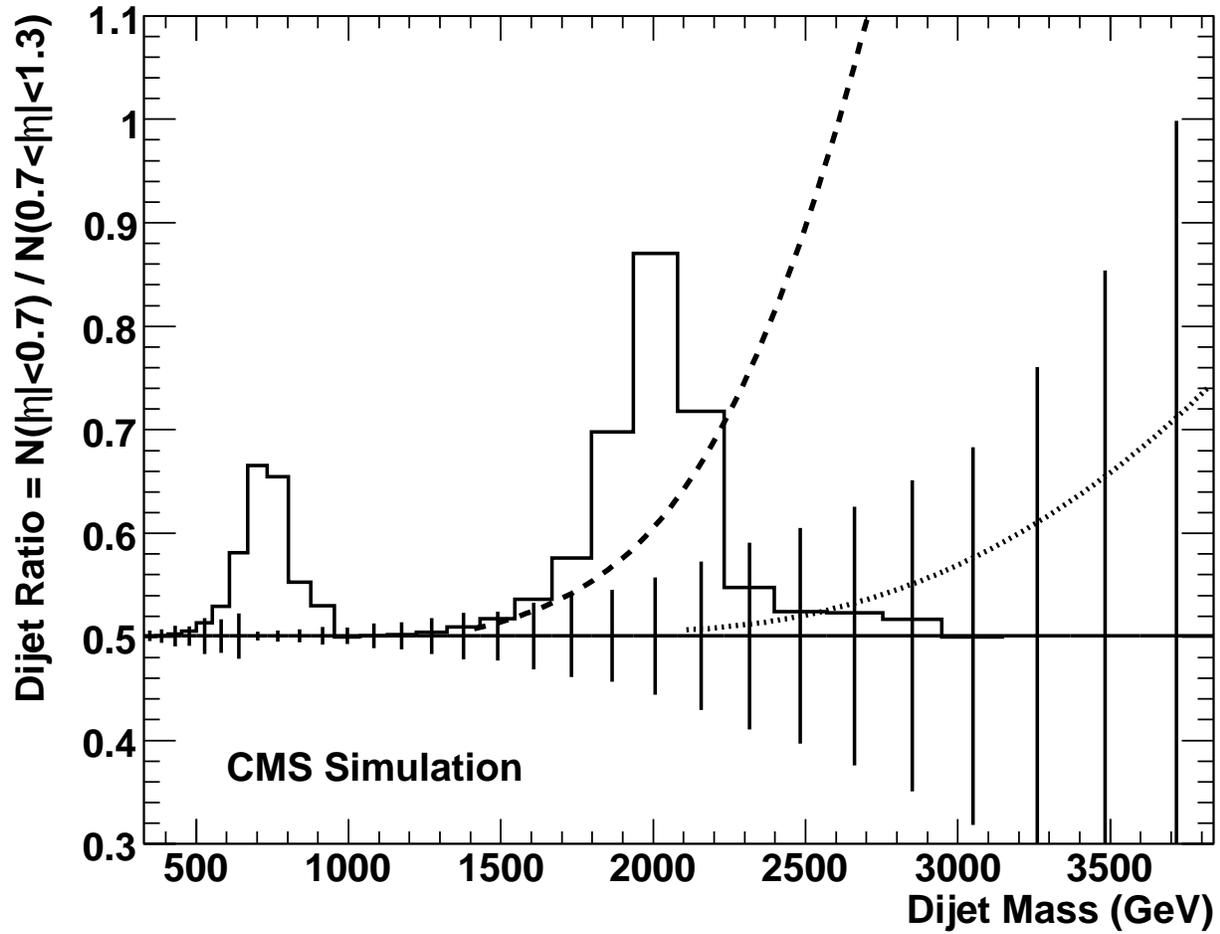}}
    \caption{ 
The dijet ratio for corrected jets expected from QCD (horizontal line), 
with statistical uncertainties (vertical bars) for an integrated luminosity of 100 pb$^{-1}$, is compared to
QCD $+$ contact interaction signals with a scale $\Lambda^+=5$ TeV (dashed) and 10 TeV (dotted), as well as to 
QCD $+$ dijet resonance signals
(histogram) with $q$* masses 
of 0.7 and 2~TeV.}
    \label{fig4}
  \end{center}
\end{figure}

CMS plans to search for both contact interactions and dijet resonances using the dijet ratio,
$r=N(|\eta|<0.7)/N(0.7<|\eta|<1.3)$, 
where $N$ is the number of events with both jets in the specified $|\eta|$ region.
The dijet ratio is sensitive to the dijet angular distribution. 
For the QCD processes, the dijet ratio is the same for corrected jets and generated jets, and
is constant at $r=0.5$ for dijet masses up to 6 TeV~\cite{refPAS}. 
Figure~\ref{fig4} shows the dijet ratio from contact interactions and dijet resonances, compared to 
the expected statistical uncertainty on the QCD processes, for 100 pb$^{-1}$ of integrated luminosity, 
including trigger prescaling. The signal from a contact interaction with scale $\Lambda^{+}=5$ TeV rises well above the 
QCD statistical errors at high dijet mass. Systematic uncertainties in the dijet 
ratio are expected to be small, since they predominantly cancel in the ratio 
as previously reported~\cite{refPTDR2,refPTDRratio}.  Using the dijet ratio, CMS can discover a 
contact interaction at scale 
$\Lambda^{+}=4$, 7 and 10 TeV with integrated luminosities of 10, 100, and 1000 pb$^{-1}$,
respectively~\cite{refPAS}.
The signal from a 2 TeV spin 1/2 $q$* produces a convincing peak in the dijet ratio, because
it has a significant rate and a relatively isotropic angular distribution compared to the 
QCD $t$-channel processes.  Fixing the cross section of the 2 TeV dijet resonance for 
$|\eta|<1.3$ at $13.6$ pb (from the $q$* model), the dijet ratio in the presence of QCD 
background increases by approximately 6\% 
when considering a spin 2 resonance decaying to both $q\bar{q}$ and $gg$ (such as a 
Randall-Sundrum graviton~\cite{refRSG}), and the dijet ratio decreases by approximately 4\% 
when considering a spin 1 resonance decaying to $q\bar{q}$ (such as a Z$^{\prime}$, axigluon,
or coloron)~\cite{refPAS}.  Hence, the sensitivity to a 2 TeV dijet resonance 
depends only weakly on the spin of the resonance. To measure the spin, we need both the dijet 
ratio and an independent measurement of the cross section of the resonance, for example, 
from the dijet mass differential cross section. Nevertheless, with sufficient luminosity, this 
simple measure of the dijet angular distribution, or a more complete evaluation of the angular
distribution, can be used to see these small variations and infer the spin of an observed 
dijet resonance.

In conclusion, CMS plans to use measurements of rate as a function of jet $p_T$ and dijet mass,
as well as a ratio of dijet rates in different $\eta$ regions, to search for new physics in 
the data sample collected during early LHC running.  With integrated luminosity samples in the range 
10--100 pb$^{-1}$, CMS will be sensitive to contact interactions and dijet resonances beyond those 
currently excluded by the Tevatron.

\ack

We gratefully acknowledge the assistance of our colleagues at CMS in preparing 
this paper.  For their constructive comments and guidance, we would like to 
thank our analysis review committee: Richard Cavanaugh, Dan Green and Luc Pape.
We appreciate the many useful comments on prose and style provided by Carlos 
Lourenco and the CMS publications committee. Finally, we would like to thank 
Gigi Rolandi for his encouragement and for leading the collaboration 
wide review of this paper, which served as a test of the publication 
process at CMS.

\clearpage

\end{document}